# Efficient Large-scale Trace Checking Using MapReduce[*]

Marcello M. Bersani[1], Domenico Bianculli[2], Carlo Ghezzi[1], Srđan Krstić[1] and Pierluigi San Pietro[1]
[1]DEEPSE group - DEIB - Politecnico di Milano, Milano, Italy
[2]SnT Centre - University of Luxembourg, Luxembourg, Luxembourg
marcellomaria.bersani@polimi.it, domenico.bianculli@uni.lu, carlo.ghezzi@polimi.it,
srdan.krstic@polimi.it, pierluigi.sanpietro@polimi.it

## ABSTRACT

The problem of checking a logged event trace against a temporal logic specification arises in many practical cases. Unfortunately, known algorithms for an expressive logic like MTL (Metric Temporal Logic) do not scale with respect to two crucial dimensions: the length of the trace and the size of the time interval for which logged events must be buffered to check satisfaction of the specification. The former issue can be addressed by distributed and parallel trace checking algorithms that can take advantage of modern cloud computing and programming frameworks like MapReduce. Still, the latter issue remains open with current state-of-the-art approaches.

In this paper we address this memory scalability issue by proposing a new semantics for MTL, called *lazy* semantics. This semantics can evaluate temporal formulae and boolean combinations of temporal-only formulae at any arbitrary time instant. We prove that *lazy* semantics is more expressive than standard point-based semantics and that it can be used as a basis for a correct parametric decomposition of any MTL formula into an equivalent one with smaller, bounded time intervals. We use *lazy* semantics to extend our previous distributed trace checking algorithm for MTL. We evaluate the proposed algorithm in terms of memory scalability and time/memory tradeoffs.

## 1. INTRODUCTION

Software systems have become more complex, distributed, and increasingly reliant on third-party functionality. The dynamic behavior of such systems makes traditional design-time verification approaches unfeasible, because they cannot analyze all the behaviors that can emerge at run time. For this reason, techniques like run-time verification and trace checking have become viable alternative for the verification of modern systems. While run-time verification checks the behavior of a system *during* its execution, trace checking is a *post-mortem* technique. In other words, to perform trace checking one must first collect and store relevant execution data (called execution traces or logs) produced by the system and then check them *offline* against the system specifications. For example, this activity is often done to inspect server logs, crash reports, and test traces, in order to analyze problems encountered at run time. More precisely, trace checking[1] is an automatic procedure for evaluating a formal specification over a trace of recorded events produced by a system. The output of the procedure is called *verdict* and states whether the system's behavior conforms to its formal specification.

The volume of the execution traces gathered for modern systems increases continuously as the systems become more and more complex. For example, a hourly page traffic statistics for Wikipedia articles collected over a period of 7 months amounts to 320GB of data [25]. This huge volume of trace data challenges the scalability of current trace checking tools [7, 15, 17, 23, 24], which are centralized and use sequential algorithms to process the trace. One possible way to efficiently perform trace checking over large traces is to use a distributed and parallel algorithm, as done in [3, 5] and also in the previous work [9] of some of the authors. These approaches rely on the MapReduce framework [13] to handle the processing of large traces. MapReduce is a programming model and an underlying execution framework for parallel and distributed processing of large quantities of data stored on a cluster of different interconnected machines (or nodes). In [9] we proposed a MapReduce algorithm that checks very large execution traces against formal specifications expressed in metric temporal logic (MTL); the algorithm exploits the structure of the formula to parallelize its evaluation.

MTL [18] is a class of temporal logic used for the specification and verification of real-time systems. It extends the well known "*Until*" temporal operator of the classical LTL with an interval that indicates the time distance within which the formula must hold. For example, property *"It is always true when a student accesses a homework assignment, he/she can provide or modify the answer within a week before a professor revokes the access."* is expressed as:

G(access → (can_write ∨ can_modify)U$_{(0, 604\,800\,000]}$revoke)

where operator U (called "*Until*") states that its right operand, the revoke event, must occur within a week (i.e., 604 800 000ms, assuming a millisecond granularity in the log) from the moment of access (expressed by the access event). It also states that the left operand must be continuously true until that happens. Operator G (called "*Globally*") simply states that property holds over the whole trace. In this logic, time can be expressed either as integer or real time-stamps, corresponding, respectively, to its discrete-time and continuous-time variants. MTL specifications may express properties that refer to different parts of the trace or to large portions of the trace at once by using large time intervals. In the example above, to check if the "*Until*" subformula holds in a single position of the trace, the algorithm needs to consider a portion of the trace corresponding, in the worst case, to a whole week of logged data. For the whole formula, this process needs to be performed for every position in the trace because of the outer "*Globally*" operator. More generally, trace checking algorithms scan a trace and typically buffer the events that satisfy the temporal constraints of the formula. The buffer is incrementally updated as the trace is scanned and the algorithms incrementally provide verdicts for the positions for which they have enough information (to determine the verdict). In [24] the authors state that the lower-bound for memory complexity of trace checking algorithms is exponential in the numeric constants occurring in the MTL formula encoded in binary. Therefore the strategy of buffering events creates a memory scalability issue for trace checking algorithms. This issue also affects distributed and parallel solutions, including our previous work [9]. More specifically, the memory scalability of a trace algorithm on a single cluster node depends exponentially on the numeric constants defining the bounds of the time intervals in the MTL formula to be checked.

---

[*]This work has been partially supported by the National Research Fund, Luxembourg (FNR/P10/03).

[1]Also called *trace validation* [22] or *history checking* [16].

The goal of this paper is to address this memory scalability issue by proposing a trace checking algorithm that exploits a new semantics for MTL, called *lazy* semantics. Unlike traditional *point-based* semantics [18], our *lazy* semantics can evaluate temporal formulae and boolean combinations of temporal-only formulae *at any arbitrary* time instant, while it evaluates atomic propositions only at time-stamped positions of the trace. We propose *lazy* semantics because it possesses certain properties that allow us to decompose any MTL formula into an equivalent MTL formula where the upper bound of all time intervals of its temporal operators is limited by some constant. This decomposition plays a major role in the context of (distributed) trace checking of formulae with large time intervals. In practice, if we want to check a formula with a large time interval, applying the decomposition entails an an equivalent formula, with smaller time intervals. We can then use our new trace checking algorithm that applies *lazy* semantics and checks the new formula in a more memory-efficient way.

We show that our proposed semantics does not hinder the expressive power of MTL: in fact we prove that MTL interpreted over *lazy* semantics is *strictly* more expressive than MTL interpreted over point-based semantics. In other words, any MTL formula interpreted over point-based semantics can be rewritten using an MTL formula interpreted over *lazy* semantics. Moreover, there are MTL formulae interpreted over *lazy* semantics that do not have an equivalent formula that can be interpreted over point-based semantics. We have integrated lazy semantics and the modified distributed trace checking algorithm into our MTLMAPREDUCE tool [19], implemented using the Apache Spark framework. The evaluation shows that the proposed approach can be used to check formulae that use very large time intervals, on very large traces, while keeping a low memory footprint. This footprint is compatible with the available configuration of common cloud instances. Moreover, our tool performs better, in terms of memory scalability, than state-of-the-art tools. We have also assessed the time and memory tradeoffs of the algorithm when different decomposition parameters are used.

In summary, the specific contributions of this paper are:
1) A new semantics for MTL, called *lazy semantics*; we prove that it is strictly more expressive than point-based semantics. 2) A parametric decomposition of MTL formulae into MTL formulae where the upper bound of all time intervals is limited by some constant; 3) A new trace checking algorithm that exploits *lazy* semantics and parametric decomposition, to check MTL formulae in a memory-efficient way; 4) The evaluation of the proposed algorithm in terms of memory scalability and time/memory tradeoffs.

The rest of the paper is structured as follows. Section 2 briefly introduces MTL interpreted over point-based semantics and the MapReduce programming model. Section 3 overviews our approach and motivates the need for *lazy* semantics and the parametric decomposition of MTL formulae. *Lazy* semantics is introduced in Section 4. Section 5 details the parametric decomposition of MTL formulae. Section 6 introduces our distributed trace checking algorithm that supports *lazy* semantics. Section 7 reports on the evaluation of our implementation. Section 8 surveys related work, while Section 9 concludes the paper.

## 2. PRELIMINARIES

## 2.1 Point-based Semantics for MTL

Let $I$ be any non-empty interval over $\mathbb{N}$ and let $\Pi$ be a finite set of atomic propositions (or *atoms*). The syntax of MTL is defined by the following grammar, where $p \in \Pi$ and $\mathsf{U}_I$ is the metric "*Until*" operator: $\phi ::= p \mid \neg\phi \mid \phi \vee \phi \mid \phi\mathsf{U}_I\phi$. Additional boolean operators and temporal operators can be derived using the usual conventions: "*Eventually*" is defined as $\mathsf{F}_I\phi \equiv \top\mathsf{U}_I\phi$; "*Globally*" is defined as $\mathsf{G}_I\phi \equiv \neg\mathsf{F}_I\neg\phi$. We adopt the convention that an interval of the form $[i, i]$ is written as "$= i$". The interval $[0, +\infty)$ in temporal operators is omitted for simplicity. We introduce the following shorthand notation: $\mathsf{F}^K(\phi) \equiv \underbrace{\mathsf{F}\mathsf{F}\ldots\mathsf{F}}_{K \text{ times}}(\phi)$, with $\mathsf{F}^0(\phi) = \phi$. Hereafter we refer to point-based semantics for MTL as MTL$_\mathsf{P}$ semantics.

**MTL$_\mathsf{P}$ semantics.** We focus on the finite-word semantics of MTL, since we apply it to the problem of trace checking. A *timed sequence* $\tau$, of length $|\tau| > 0$, is a sequence $\tau_0\tau_1\ldots\tau_{|\tau|-1}$ of values $\tau_i \in \mathbb{R}$ such that $0 < \tau_i < \tau_{i+1}$ for each $0 \leq i < |\tau|-1$, i.e., the sequence is *strictly monotonic*. A *word* $\sigma$ over the alphabet $2^\Pi$ is a sequence $\sigma_0\sigma_1\ldots\sigma_{|\sigma|-1}$ such that $\sigma_i \in 2^\Pi$ for all $0 \leq i < |\sigma|$, where $|\sigma|$ denotes the length of the word. A *timed word* [1] $\omega = \omega_0\omega_1\ldots\omega_{|\omega|-1}$ is a word over $2^\Pi \times \mathbb{R}$, i.e., a sequence of pairs $\omega_i = (\sigma_i, \tau_i)$, where $\sigma_0\ldots\sigma_{|\omega|-1}$ is a word over $2^\Pi$ and $\tau_0\ldots\tau_{|\omega|-1}$ is a timed sequence. A pair $\omega_i$ is also called an *element* of the timed word. Moreover, notice that in this definition $i$ refers to a particular *position* of the element $\omega_i$ in the timed word $\omega$, while $\tau_i$ refers to the *time instant* or *time-stamp* of the element $\omega_i$. We abuse the notation and represent a timed word equivalently as a pair containing a word and a timed sequence of the same length, i.e., $\omega = (\sigma, \tau)$. A *timed language* over $2^\Pi$ is a set of timed words over $2^\Pi$. MTL$_\mathsf{P}$ semantics on timed words is given in Figure 1, where the point-based satisfaction relation $\models_P$ is defined with respect to a timed word $(\sigma, \tau)$, a position $i \in \mathbb{N}$, and MTL formulae $\phi$ and $\psi$. Note that, due to the strictly monotonic definition of the timed sequence $\tau$, the metric "*Next*" operator can be defined as $\mathsf{X}_I\phi \equiv \bot\mathsf{U}_{I-\{0\}}\phi$. $L_P(\phi)$ is a timed language defined by a formula $\phi$ when interpreted over the MTL$_\mathsf{P}$ semantics, i.e., $L_P(\phi) = \{(\sigma, \tau) \mid (\sigma, \tau, 0) \models_P \phi\}$

$$(\sigma, \tau, i) \models_P p \text{ iff } p \in \sigma_i \text{ for } p \in \Pi$$
$$(\sigma, \tau, i) \models_P \neg \phi \text{ iff } (\sigma, \tau, i) \not\models_P \phi$$
$$(\sigma, \tau, i) \models_P \phi \vee \psi \text{ iff } (\sigma, \tau, i) \models_P \phi \text{ or } (\sigma, \tau, i) \models_P \psi$$
$$(\sigma, \tau, i) \models_P \phi \mathsf{U}_I \psi \text{ iff } \exists j.(i \leq j < |\sigma| \text{ and } \tau_j - \tau_i \in I \text{ and}$$
$$(\sigma, \tau, j) \models_P \psi \text{ and } \forall k.(i < k < j \text{ then } (\sigma, \tau, k) \models_P \phi)$$

**Figure 1:** $\mathsf{MTL_P}$ semantics on timed words.

## 2.2 The MapReduce programming model

MapReduce [13] is a programming model, developed by Google, for processing and analyzing large data sets using a parallel, distributed infrastructure. The MapReduce programming model uses two user-defined functions, *map* and *reduce*, that are inspired by the homonymous functions that are typically found in functional programming languages. The *map* function receives a key-value pair associated with the input data and returns a set of intermediate key-value pairs; its signature is `map(k:K`$_1$`,v:V`$_1$`):list[(k:K`$_2$`, v:V`$_2$`)]`. The *reduce* function is applied to all the intermediate values that have the same intermediate key in order to combine the derived data appropriately; its signature is `reduce(k:K`$_2$`, list(v:V`$_2$`)):list[v:V`$_2$`]`. In the definitions above, $K_1$ and $K_2$ are types for keys and $V_1$ and $V_2$ are types for values.

Besides the actual programming model, MapReduce is also a framework that provides, in a transparent way to developers, parallelization, fault tolerance, locality optimization, and load balancing. The MapReduce framework is responsible for partitioning the input data, scheduling and executing the *Map* and *Reduce* tasks (also called *mappers* and *reducers*, respectively) on a cluster of available nodes, and for managing communication and data transfer (usually leveraging a distributed file system). More in detail, the execution of a MapReduce operation (called *job*) proceeds as follows. First, the system splits the input into blocks[2] of a certain size using an *InputReader*, generating input key/value pairs. It then assigns each input block to mappers, which are processed in parallel by the nodes in the distributed architecture. A mapper reads the corresponding input block and passes the set of key/value pairs to the *map* function, which generates a set of intermediate key/value pairs. Notice that each run of the *map* function is stateless, i.e., the transformation of a single key/value pair does not depend on any other key/value pair. The next phase is called *shuffle and sort*: it takes the intermediate data generated by each mapper, sorts them based on the intermediate data generated from the other nodes, divides these data into regions to be processed by reducers, and distributes these data on the nodes where the reducers will be executed. The division of intermediate data into regions is done by a *partitioning function*, which depends on the (user-specified) number of reducers and the key of the intermediate data. Each reducer executes the *reduce* function, which produces the output data. This output is appended to a final output file for this reduce partition. The output of the MapReduce job will then be available in several files, one for each used reducer. Multiple MapReduce calls can be linked together in sequence to perform complex data processing.

---

[2]Also called input splits or chunks.

| Atoms: | $\{p\}$ | $\{p\}$ | | $\{q\}$ | | $\{p,q\}$ | | $\{p,q\}$ | $\{q\}$ | $\{q\}$ |
|---|---|---|---|---|---|---|---|---|---|---|
| Time-stamps: | 1 | 2 | 3 | 4 | 5 | 6 | 7 | 8 | 9 | 10 |
| $p$ | ⊤ | ⊤ | | ⊥ | | ⊤ | | ⊤ | ⊥ | ⊥ |
| $\mathsf{F}_{[3,7]}(p)$ | (⊤) | ⊤ | | ⊤ | | ⊥ | | ⊥ | ⊥ | ⊥ |

**Figure 2:** Evaluation of formula $\Phi = \mathsf{F}_{[3,7]}(p)$.

## 3. MOTIVATION AND OVERVIEW OF THE APPROACH

As mentioned in Section 1, trace checking is an automatic procedure for evaluating a formal specification over a trace of recorded events produced by a system. Since traces can be seen as a sequence of time-stamped elements (where each element records one or more events), we use timed words as abstract models of traces. Hence, a pair $\omega_i = (\sigma_i, \tau_i)$ corresponds to the $i$-th element of the trace, where $\sigma_i$ represents all the event(s) with time-stamp $\tau_i$.

Trace checking algorithms handle metric temporal operators by buffering elements of the trace. The time interval specified in the metric temporal formula to check determines the portion of the trace that needs to be considered to decide whether the formula is true in a single position of the trace. Depending on the particular MTL formula that is being checked, in the worst case this process needs to be repeated for every position in the trace[3]. What trace checking algorithms typically do is to keep the relevant portion of the trace in a buffer as they scan the trace. The buffer is updated incrementally while the algorithm scans and produces verdicts for the following elements in the trace. The procedure for updating the buffer consists of adding a newly-scanned element $e$ of the trace and removing the elements whose time-stamps do not satisfy the temporal constraint of the formula to check, when evaluated with respect to the time-stamp of $e$. Buffering elements presents a memory scalability issue if a metric temporal formula with a large interval needs to be processed. Let us present an example to motivate the need for *lazy* semantics.

EXAMPLE 1. *Consider formula $\Phi = \mathsf{F}_{[3,7]}(p)$ and its evaluation on the following trace (represented as a timed word):* $(\{p\},1)$, $(\{p\},2)$, $(\{q\},4)$, $(\{p,q\},6)$, $(\{p,q\},8)$, $(\{q\},9)$, $(\{q\},10)$. *The timed word, shown in Figure 2, is defined over the set of atoms $\Pi = \{p,q\}$; its length is 7 and it spans over 10 time units. The first two rows in the picture represent its atoms and time-stamps; the last two rows show, respectively, the evaluation of subformula $p$ and formula $\mathsf{F}_{[3,7]}(p)$ using point-based semantics. As shown in the last row of Figure 2, according to point-based semantics, formula $\mathsf{F}_{[3,7]}(p)$ holds at positions 1, 2 and 3.*

For a formula of the form $\mathsf{F}_{[a,b]}(p)$, the algorithm needs to buffer, in the worst case (i.e., in case there exists an element in correspondence of every time instant), at most $b + 1$ elements. For example, to evaluate formula $\mathsf{F}_{[3,7]}(p)$ at time instant 2, in the worst case the algorithm will buffer 8 elements, i.e., all the elements whose time-stamp ranges from 2 to 9. The elements with time-stamps ranging from 6 to 9 satisfy the time interval constraint of the formula; the others are kept for the evaluation of the formula at subsequent positions. Let us assume that the execution infrastructure could only store 5 elements in the buffer, for example because of limited memory. The worst-case requirement of keeping 8 elements in the buffer would then be too demanding for the infrastructure, in terms of memory scalability. To lower the memory requirement for the buffer we would need a formula with a smaller time interval and expressing the same property as $\Phi$. In other words, one might ask whether there is an MTL formula equivalent to $\Phi$ with all the intervals bounded by the constant 4 (and thus requiring to store at most 4+1=5 elements in the buffer).

---

[3]For example, if a "*Globally*" temporal operator is used.

Let us consider formula $\Phi' = \mathsf{F}_{[3,4]}(p) \vee \mathsf{F}_{[4,4]}(\mathsf{F}_{[0,3]}(p))$: a naïve, intuitive interpretation might lead us to think that it defines the same property as $\Phi$. Roughly speaking, instead of checking if $p$ eventually occurs within the entire $[3,7]$ time interval, $\Phi'$ checks if $p$ either occurs in the $[3,4]$ interval (as specified by subformula $\mathsf{F}_{[3,4]}(p)$) or in the interval $[0,3]$ when evaluated exactly 4 time instants in the future (as specified by subformula $\mathsf{F}_{[4,4]}(\mathsf{F}_{[0,3]}(p))$). Figure 3 shows the evaluation of formula $\Phi'$ over the same trace used in Figure 2. As one can see, formula $\Phi'$ does not have the same evaluation as $\Phi$ on the same trace. More specifically, at time instant 1 $\Phi'$ is false while $\Phi$ is true (see the values circled in both figures). By analyzing the evaluation of $\Phi'$, one can notice that subformula $\mathsf{F}_{[4,4]}(\mathsf{F}_{[0,3]}(p))$ at time instant 1 refers to the value of $\mathsf{F}_{[0,3]}(p)$ at time instant 5, which does not have a corresponding element in the trace. If there was an element at time instant 5, $\mathsf{F}_{[0,3]}(p)$ would be true since $p$ holds at instant 6.

The above example shows that the evaluation of temporal subformulae according to point-based semantics depends on the existence of certain elements in the trace. It also shows that point-based semantics is not suitable to support the intuitive decomposition of MTL formulae into equivalent ones with smaller time intervals, like the one from $\Phi$ to $\Phi'$ shown above. We maintain that this constitutes a limitation for the application of point-based semantics in the context of trace checking. Therefore, in this paper we propose a new, alternative semantics for MTL, called *lazy* semantics.

The main feature of *lazy* semantics is that it evaluates temporal formulae and boolean combinations of temporal-only formulae *at any arbitrary* time instant, regardless of the existence of the corresponding elements in the trace. The existence of the elements is only required when evaluating atoms. This features allows us to decompose any MTL formula into an equivalent MTL formula in which the upper bound of all time intervals of its temporal operators is limited by some constant. Such a decomposition can be used as a pre-processing step of a trace checking algorithm, which can then perform in a more memory-efficient way.

In the following sections we first introduce *lazy* semantics (Section 4) and formalize the notion of the decomposition

| Atoms: | $\{p\}$ | $\{p\}$ | | $\{q\}$ | | $\{p,q\}$ | | $\{p,q\}$ | $\{q\}$ | $\{q\}$ |
|---|---|---|---|---|---|---|---|---|---|---|
| Time-Stamps: | 1 | 2 | 3 | 4 | 5 | 6 | 7 | 8 | 9 | 10 |
| $p$ | ⊤ | ⊤ | | ⊥ | | ⊤ | | ⊤ | ⊥ | ⊥ |
| $\mathsf{F}_{[3,4]}(p)$ | ⊥ | ⊤ | | ⊤ | | ⊥ | | ⊥ | ⊥ | ⊥ |
| $\mathsf{F}_{[0,3]}(p)$ | ⊤ | ⊤ | | ⊤ | | ⊤ | | ⊤ | ⊥ | ⊥ |
| $\mathsf{F}_{[4,4]}(\mathsf{F}_{[0,3]}(p))$ | ⊥ | ⊤ | | ⊤ | | ⊥ | | ⊥ | ⊥ | ⊥ |
| $\Phi'$ | (⊥) | ⊤ | | ⊤ | | ⊥ | | ⊥ | ⊥ | ⊥ |

**Figure 3: Evaluation of formula $\Phi' = \mathsf{F}_{[3,4]}(p) \vee \mathsf{F}_{[4,4]}(\mathsf{F}_{[0,3]}(p))$.**

$$(\sigma, \tau, t) \models_L p \text{ iff } \exists i.(0 \leq i < |\sigma| \text{ and } t = \tau_i \text{ and } p \in \sigma_i)$$
$$(\sigma, \tau, t) \models_L \neg \phi \text{ iff } (\sigma, \tau, t) \not\models_L \phi$$
$$(\sigma, \tau, t) \models_L \phi \vee \psi \text{ iff } (\sigma, \tau, t) \models_L \phi \text{ or } (\sigma, \tau, t) \models_L \psi$$
$$(\sigma, \tau, t) \models_L \phi \mathsf{U}_I \psi \text{ iff } \exists t'.(t' \geq t \text{ and } t' - t \in I \text{ and }$$
$$(\sigma, \tau, t') \models_L \psi \text{ and } \forall t''.(t < t'' < t' \text{ and } \exists i.(0 \leq i < |\sigma| \text{ and }$$
$$t'' = \tau_i) \text{ then } (\sigma, \tau, t'') \models_L \phi))$$

**Figure 4: MTL$_L$ semantics on timed words.**

exemplified above (Section 5). Afterwards, in Section 6 we describe the modifications to our previous trace checking algorithm [9], required to preprocess the formula and support *lazy* semantics.

## 4. LAZY SEMANTICS FOR MTL

The following example shows an anomalous case of MTL$_P$ semantics that lazy semantics for MTL (denoted as MTL$_L$ semantics) intends to remedy. Consider a timed word $w = (\sigma, \tau) = (\{q\}, 1)(\{p\}, 7) \ldots$ and two MTL formulae $\psi_1 : \mathsf{F}_{=6}p$ and $\psi_2 : \mathsf{F}_{=3}\mathsf{F}_{=3}p$. The intuitive meaning of the two formulae is the same: $p$ holds 6 time units after the origin, i.e., at time-stamp 7. However, when evaluated in the first position of $w$ using the MTL$_P$ semantics, the two formulae have opposite values: $\psi_1$ correctly evaluates to true, but $\psi_2$ to false, since in $\psi_2$ the outermost $\mathsf{F}_{=3}$ subformula is trivially false, because there is no position that is exactly 3 time instants in the future with respect to the origin. The two formulae, instead, are equivalent under the MTL$_L$ semantics, where they both evaluate to true. Indeed, this is true also over signal-based semantics [11]; however, signals are not very practical for monitoring and trace checking, which usually operate on logs that are best modeled as a sequence of discrete time-stamped observations, i.e., timed words.

**MTL$_L$ semantics.** MTL$_L$ semantics on timed words is given in Figure 4, in terms of the satisfaction relation $\models_L$, with respect to a timed word $(\sigma, \tau) = (\sigma_0, \tau_0)(\sigma_1, \tau_1) \ldots$ and a time instant $t \in \mathbb{R}^+$; $p$ is an atom and $\phi$ and $\psi$ are MTL formulae. An MTL formula $\phi$, when interpreted over MTL$_L$ semantics, defines a timed language $L_L(\phi) = \{(\sigma, \tau) | (\sigma, \tau, 0) \models_L \phi\}$. The main difference between MTL$_P$ and MTL$_L$ semantics is that MTL$_P$ evaluates formulae only at positions $i$ of a timed word, while MTL$_L$ inherits a feature of signal-based semantics, namely it may evaluate (non-atomic) formulae at any possible time instant $t$, even if there is no time-stamp equal to $t$. For example, according to the MTL$_P$ semantics, an "*Until*" formula $\phi \equiv \psi_1 \mathsf{U}_I \psi_2$ evaluates to false in case there are no positions in the interval $I$, due to the existential quantification on $j$ (see Figure 1). Conversely, under the MTL$_L$ semantics, the evaluation of $\phi$ depends on the evaluation of $\psi_2$. If the latter is an atom then formula $\phi$ also evaluates to false, because of the existential quantifier in the MTL$_L$ semantics of atoms. However, if $\psi_2$ is a temporal formula or a boolean combination of temporal-only formulae (e.g., other "*Until*" formulae), it will be evaluated in the part of the timed word that satisfies the interval of $\phi$. Hereafter we refer to the MTL formulae interpreted over the MTL$_L$ semantics as "MTL$_L$ formulae"; similarly, "MTL$_P$ formulae" are MTL formulae interpreted over the MTL$_P$ semantics.

Let $\mathbb{M}(\Pi)$ be the set of all formulae that can be derived from the MTL grammar shown in Section 2.1, using $\Pi$ as the set of atoms. We show that any language $L_P(\phi)$ defined using some $\text{MTL}_P$ formula $\phi$ can be defined using an $\text{MTL}_L$ formula obtained after applying the translation $l2p$ : $\mathbb{M}(\Pi) \to \mathbb{M}(\Pi)$ to $\phi$, i.e., $L_p(\phi) = L_L(l2p(\phi))$ for any $\phi$. The $l2p$ translation is defined as follows:

$$l2p(p) \equiv p, p \in \Pi; \quad l2p(\phi \vee \psi) \equiv l2p(\phi) \vee l2p(\psi)$$
$$l2p(\neg \phi) \equiv \neg l2p(\phi); \quad l2p(\phi \mathsf{U}_I \psi) \equiv l2p(\phi) \mathsf{U}_I (\varphi_{act} \wedge l2p(\psi))$$

where $\varphi_{act} \equiv a \vee \neg a$ for some $a \in \Pi$.

The goal of $l2p$ is to prevent the occurrence of the nesting of temporal operators, i.e., to avoid the presence of (sub)formulae like $\mathsf{F}_{=3}\mathsf{F}_{=3}p$. As discussed above in the example, nested temporal operators are interpreted differently over the two semantics. Nesting is avoided by rewriting the right argument of every "Until" (i.e., the "existential" component of "Until"). The argument is conjuncted with a formula $\varphi_{act}$ that evaluates to true (under both semantics) if there exists a position in the underlying timed word; otherwise $\varphi_{act}$ evaluates to false. To explain this intuition, let us evaluate $\varphi_{act}$ over a timed word $(\sigma, \tau)$ over the alphabet $\Pi = \{a\}$. Under point-based semantics, $(\sigma, \tau, i) \models_P \varphi_{act} \equiv (\sigma, \tau, i) \models_P a \vee \neg a$ is true for any position $i$, since either $a$ belongs to $\sigma_i$ or not. However, the same does not hold for lazy semantics. According to lazy semantics, $(\sigma, \tau, t) \models_L \varphi_{act}$ is true only in those time instants $t$ for which there exists $i$ such that $\tau_i = t$ and therefore exists the corresponding $\sigma_i$ (to which $a$ can belong or not).

LEMMA 1. *Given an MTL formula $\phi$ and a timed word $\omega = (\sigma, \tau)$, for any $i \geq 0$, the following equivalence (modulo l2p translation) holds: $(\sigma, \tau, i) \models_P \phi$ iff $(\sigma, \tau, \tau_i) \models_L l2p(\phi)$.*

The proof of the lemma is in the appendix A.

THEOREM 1. *Any timed language defined by an $\text{MTL}_P$ formula can be defined by an $\text{MTL}_L$ formula over the same alphabet.*

PROOF. By Lemma 1, for $i = 0$. □

Notice that the translation $l2p$ defines a syntactic MTL fragment where temporal or boolean combination of temporal-only operators cannot be nested. In this fragment $\text{MTL}_P$ and $\text{MTL}_L$ formulae define the same languages. However, if we consider the complete definition of MTL, without syntactic restrictions, the class of timed languages defined by $\text{MTL}_L$ formulae strictly includes the class of languages defined by $\text{MTL}_P$ formulae. In other words, MTL interpreted over lazy semantics is *strictly* more expressive than MTL interpreted over point-based semantics; this result is established by the following theorem.

THEOREM 2. *There exists a timed language defined by some $\text{MTL}_L$ formula that cannot be defined by any $\text{MTL}_P$ formula.*

PROOF. Consider the language of timed words $\{(\sigma, \tau) : \exists i \exists j (i \leq j \wedge (\sigma, \tau, i) \models_L b \wedge (\sigma, \tau, j) \models_L c \wedge \tau_j \leq 2)\}$. It is defined by the $\text{MTL}_L$ formula $\Phi = \Phi_1 \vee \Phi_2 \vee \Phi_3$, where $\Phi_1 = (\mathsf{F}_{(0,1)}b) \wedge (\mathsf{F}_{[1,2]}c) \vee (\mathsf{F}_{(0,1)}b) \wedge (\mathsf{F}_{(1,2]}c)$; $\Phi_2 = \mathsf{F}_{(0,1]}(b \wedge \mathsf{F}_{(0,1]}c)$ and $\Phi_3 = \mathsf{F}_{(0,1]}((\mathsf{F}_{(0,1)}b) \wedge (\mathsf{F}_{[1,1]}c))$. Formula $\Phi$ cannot be represented by any $\text{MTL}_P$ formula (see reference [11], prop. 6). □

## 5. PARAMETRIC DECOMPOSITION

In this section we show that *lazy* semantics allows for a parametric decomposition of MTL formulae into MTL formulae where the upper bound of all intervals of the temporal operators is limited by some constant $K$ (the parameter of the decomposition). This structural characteristic will then be used in the trace checking algorithm presented thereafter.

We first introduce some notation and show some properties of lazy semantics that will be used to prove the correctness of the decomposition. We define the operator $\oplus$ over intervals in $\mathbb{N}$ such that $I \oplus J = \{i + j \mid \forall i \in I \text{ and } j \in J\}$.

LEMMA 2. *For any timed word $(\sigma, \tau)$ and $t \geq 0$,*

$$(\sigma, \tau, t) \models_L \mathsf{F}_I \mathsf{F}_J \phi \text{ iff } (\sigma, \tau, t) \models_L \mathsf{F}_{I \oplus J} \phi.$$

COROLLARY 1. *For any timed word $(\sigma, \tau)$ and $t, N \geq 0$,*

$$(\sigma, \tau, t) \models_L \mathsf{F}^N_{=K} \phi \text{ iff } (\sigma, \tau, t) \models_L \mathsf{F}_{=K \cdot N}.$$

LEMMA 3. *For any timed word $(\sigma, \tau)$ and $t \geq 0$,*

$$(\sigma, \tau, t) \models_L \mathsf{F}_I \phi \vee \mathsf{F}_J \phi \text{ iff } (\sigma, \tau, t) \models_L \mathsf{F}_{I \cup J} \phi, \text{ if } I \cap J \neq \emptyset.$$

The proof of the above corollary and lemmata is in the appendix A.

Hereafter, we focus on bounded MTL formulae, i.e., formulae where intervals are always finite. We present the parametric decomposition by referring to the bounded "Eventually" operator. The bounded "Until" and "Globally" operators can be expressed in terms of the bounded "Eventually" operator using the usual equivalences ; moreover, we remark that the decomposition does not affect atoms and is applied recursively to boolean operators. We use angle brackets (symbols "⟨" and "⟩") in the definition of the decomposition to cover all four possible cases of open (denoted with round brackets) and closed (denoted with square brackets) intervals; the definition is valid for any instantiation of the symbols as long as they are consistently replaced on the right-hand side.

The *decomposition* $\mathcal{L}_K$ of MTL formulae with respect to *parameter* $K$ is the translation $\mathcal{L}_K : \mathbb{M}(\Pi) \to \mathbb{M}(\Pi)$ such that $\mathcal{L}_K(\mathsf{F}_{\langle a,b \rangle}\phi) =$

$$\begin{cases} \mathsf{F}_{\langle a,b \rangle} \mathcal{L}_K(\phi) & , b \leq K \\ \mathsf{F}^{\lfloor \frac{a}{K} \rfloor}_{=K} (\mathsf{F}_{\langle a \bmod K, b - \lfloor \frac{a}{K} \rfloor \cdot K \rangle} \mathcal{L}_K(\phi)) & , K < b \leq \lfloor \frac{a}{K} + 1 \rfloor \cdot K \\ \mathsf{F}^{\lfloor \frac{a}{K} \rfloor}_{=K} (\mathsf{F}_{\langle a \bmod K, K]} \mathcal{L}_K(\phi) \vee & , b > \lfloor \frac{a}{K} + 1 \rfloor \cdot K \\ \quad \mathsf{F}^{\lfloor \frac{a}{K} \rfloor}_{=K} (\mathsf{F}_{=K}(\mathcal{D}_F(\mathcal{L}_K(\phi), K, b - \lfloor \frac{a}{K} + 1 \rfloor \cdot K))) \end{cases}$$

where

$$\mathcal{D}_F(\psi, K, h) = \begin{cases} \mathsf{F}_{[0,h)}\psi & , h \leq K \\ \mathsf{F}_{[0,K]}\psi \vee \mathsf{F}_{=K}(\mathcal{D}_F(\psi, K, h - K)) & , h > K \end{cases}$$

The decomposition $\mathcal{L}_K$ considers three cases depending on the values of $a$, $b$, and $K$. In the first case we have $b \leq K$, which means that the upper bound of the temporal interval $[a, b]$ in the input formula is smaller than $K$, therefore no decomposition is needed. The other two cases consider input formulae where $b > K$. The second case is characterized by $b \leq \lfloor \frac{a}{K} + 1 \rfloor \cdot K \equiv b \leq \lfloor \frac{a}{K} \rfloor \cdot K + K$. The decomposition yields a formula of the form $\mathsf{F}^{\lfloor \frac{a}{K} \rfloor}_{=K}(\alpha)$, where $\alpha = \mathsf{F}_{[a \bmod K, b - \lfloor \frac{a}{K} \rfloor \cdot K]} \mathcal{L}_K(\phi)$ is equivalent to the input formula $\mathsf{F}_{[a,b]}(\phi)$ evaluated at time instant $\lfloor \frac{a}{K} \rfloor \cdot K$. Notice that according to Corollary 1, the argument $\alpha$ in $\mathsf{F}_{=K}^{\lfloor \frac{a}{K} \rfloor}(\alpha)$ is evaluated at time instant $\lfloor \frac{a}{K} \rfloor \cdot K$. The third case is characterized by $b > \lfloor \frac{a}{K} + 1 \rfloor \cdot K$.

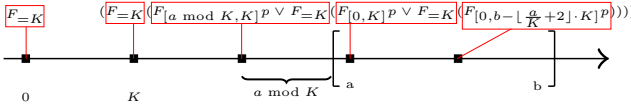

**Figure 5:** $\mathcal{L}_K$ decomposition of formula $\mathsf{F}_{[a,b]}p$.

We illustrate the decomposition of $\mathsf{F}_{[a,b]}p$ with $p \in \Pi$ by referring to the example in Figure 5, where the black squares divide the timeline into segments of length $K$. We refer to each position in the timeline pinpointed by a black square as a $K$-position. The big brackets enclose the interval $[a,b]$ relative to time instant 0. Moreover, we assume some values for $a$ and $K$ such that $\lfloor \frac{a}{K} \rfloor = 2$; hence, in the figure the position of $a$ in the timeline is between the marks corresponding to $2K$ and $3K$. The application of $\mathcal{L}_K(\mathsf{F}_{[a,b]}p)$ returns the formula $F_{=K}(F_{=K}(F_{[a \bmod K, K]}p \lor F_{=K}(F_{[0,K]}p \lor F_{=K}(F_{[0,b-\lfloor \frac{a}{K}+2\rfloor \cdot K]}p))))$, which is shown above the timeline, spanning through its length such that each subformula is written above the corresponding $K$-position where it is evaluated. Since $\lfloor \frac{a}{K} \rfloor = 2$ there are two subformulae of the form $\mathsf{F}_{=K}$ evaluated in the first two $K$-positions. Unlike the previous case, the interval $[a,b]$ is too big to allow for rewriting the input formula into another formula with a single F operator with bounded length. Hence, we use three subformulae: 1) $\mathsf{F}_{[a \bmod K]}p$ evaluated at the third $K$-position, 2) $\mathsf{F}_{[0,K]}p$ evaluated at the fourth $K$-position, and 3) $\mathsf{F}_{[0,b-\lfloor \frac{a}{K}+2\rfloor \cdot K]}p$ evaluated at the fifth $K$-position; the last two subformulae are obtained from the definition of $\mathcal{D}_F$. Notice that if $K$ is set to be equal to one, the $\mathcal{L}_K$ decomposition boils down to the reduction of MTL to LTL.

THEOREM 3. *Given an* $\mathsf{MTL_L}$ *formula* $\phi$, *a timed word* $(\sigma, \tau)$ *and a positive constant* $K$, *we have that:*

$$(\sigma, \tau, 0) \models_L \phi \text{ iff } (\sigma, \tau, 0) \models_L \mathcal{L}_K(\phi)$$

*and the upper bound of every bounded interval in all temporal subformulae of* $\mathcal{L}_K(\phi)$ *is less than or equal to* $K$.

PROOF. We can prove this statement by showing that $\mathcal{L}_K(\phi)$ can always be rewritten back as $\phi$ and vice versa using Lemmas 2 and 3. The complete proof is provided in the appendix. □

## 6. TRACE CHECKING $\mathsf{MTL_L}$ FORMULAE WITH MAPREDUCE

The theoretical results presented in Section 5 can be applied to improve the memory scalability of the distributed trace checking algorithm based on MapReduce programming model, and introduced by some of the authors in previous work [9]. Although the algorithm presented in [9] was designed to perform trace checking of properties written in SOLOIST [10] (an extension of MTL with aggregating temporal modalities), here we consider, without loss of generality (see [10]), only its MTL subset. In the rest of this section, after introducing some additional notation, we give an overview of the algorithm's execution flow, and detail the modifications (emphasized with grey boxes in Figure 6) applied to the original algorithm defined in [9] to support $\mathsf{MTL_L}$ semantics.

**Additional notation.** Let $\phi$ and $\psi$ be MTL formulae. The set of all proper subformulae of $\phi$ is denoted with $\mathsf{sub}(\phi)$; notice that for atoms $p \in \Pi$, $\mathsf{sub}(p) = \emptyset$. The size of a formula $\phi$, denoted $|\phi|$, is defined as the number of its non-proper subformulae, i.e. $|\phi| = |\mathsf{sub}(\phi)| + 1$. The set $\mathsf{sub}_a(\phi) = \{p \mid p \in \mathsf{sub}(\phi), \mathsf{sub}(p) = \emptyset\}$ is the set of atoms of formula $\phi$. The set $\mathsf{sub}_d(\phi) = \{\alpha \mid \alpha \in \mathsf{sub}(\phi), \forall \beta \in \mathsf{sub}(\phi), \alpha \notin \mathsf{sub}(\beta)\}$ is called the set of all *direct subformulae* of $\phi$; $\phi$ is called the *superformula* of all formulae in $\mathsf{sub}_d(\phi)$. The set $\mathsf{sup}_\psi(\phi) = \{\alpha \mid \alpha \in \mathsf{sub}(\psi), \phi \in \mathsf{sub}_d(\alpha)\}$ is the set of all subformulae of $\psi$ that have formula $\phi$ as *direct subformula*. The *height* $h(\phi)$ of $\phi$ is defined recursively as:

$$h(\phi) = \begin{cases} \max\{h(\psi) \mid \psi \in \mathsf{sub}_d(\phi)\} + 1 & \text{if } \phi \notin \Pi, \\ 1 & \text{otherwise.} \end{cases}$$

For example, given the formula $\gamma = \mathsf{F}_{[2,4]}(a \land b) \mathsf{U}_{(30,100)} \neg c$, we have: $\mathsf{sub}(\gamma) = \{a, b, c, a \land b, \neg c, \mathsf{F}_{[2,4]}(a \land b)\}$ is the set of all subformulae of $\gamma$; $\mathsf{sub}_a(\gamma) = \{a, b, c\}$ is the set of *atoms* in $\gamma$; $\mathsf{sub}_d(\gamma) = \{\mathsf{F}_{[2,4]}(a \land b), \neg c\}$ is the set of direct subformulae of $\gamma$; $\mathsf{sup}_\gamma(a) = \mathsf{sup}_\gamma(b) = \{a \land b\}$ shows that the sets of superformulae of $a$ and $b$ in $\gamma$ coincide; and the height of $\gamma$ is 4, since $h(a) = h(b) = h(c) = 1$, $h(\neg c) = h(a \land b) = 2$, $h(\mathsf{F}_{[2,4]}(a \land b)) = 3$ and therefore $h(\gamma) = \max\{h(\mathsf{F}_{[2,4]}(a \land b)), h(\neg c)\} + 1 = 4$.

**Overview.** The algorithm takes as input a non-empty execution trace $T$ and an MTL formula $\Phi$ and provides a verdict whether the trace satisfies the formula or not. Before the algorithm is used we assume that the execution infrastructure, i.e., the cluster of machines, is configured and running. We also assume that one can easily estimate through experimentation the largest time interval bound $K_\mathsf{cluster}$ *manageable* by the cluster, i.e., the largest bound that does not trigger memory saturation. The bound $K_\mathsf{cluster}$ depends on the memory configuration of the node in the cluster with the least amount of memory available. Once we have this information, we can preprocess the input formula $\Phi$, leveraging the theoretical results of Section 5. If the temporal operators in $\Phi$ have bounded intervals less than $K_\mathsf{cluster}$, we apply the unmodified version of the original algorithm [9], which evaluates formulae according to point-based semantics. Otherwise, we have to transform the original formula into an equivalent one that can be efficiently checked. This transformation is achieved by first interpreting the input formula $\Phi$ over lazy semantics: to preserve its meaning, we apply the $l2p$ transformation. Afterwards, given the parameter $K_\mathsf{cluster}$, we rewrite the formula using the $\mathcal{L}_{K_\mathsf{cluster}}$ decomposition (i.e., the $\mathcal{L}_K$ decomposition instantiated with parameter $K_\mathsf{cluster}$) and obtain the formula $\Phi_L^{K_\mathsf{cluster}} = \mathcal{L}_{K_\mathsf{cluster}}(l2p(\Phi))$. Thanks to Theorem 3, this formula contains intervals no greater than $K_\mathsf{cluster}$ and is equivalent to $\Phi$.

The trace is modeled as a timed word, i.e., we have $T = (\sigma, \tau)$. We call $\omega_i = (\sigma_i, \tau_i)$ an element of the trace $T$ at position $i$. It contains the set of atoms $\sigma_i \subseteq \Pi$ that hold at position $i$ and an integer time-stamp $\tau_i$. We assume that the execution trace is saved in the distributed file system of the cluster on which the distributed algorithm is executed. This is a realistic assumption since in a distributed setting it is possible to collect logs, as long as there is a total order among the time-stamp induced by some clock synchronization protocol.

The trace checking algorithm processes the trace itera-tively, through a sequence of MapReduce executions. The number of MapReduce iterations is equal to the height of the MTL formula $\Phi$. The first MapReduce iteration parses the input trace from the distributed file system, applies the `map` and `reduce` functions and passes the output (a set of tuples) to the next iteration. Each subsequent iteration receives the set of tuples from the respective previous iteration in the expected internal format, thus parsing is performed only in the first iteration. A subsequent iteration $l$ (where $1 < l \leq h(\Phi)$) receives the set of tuples from the iteration $l-1$. The set of tuples contains all the positions where the subformulae of $\Phi$ of height $l-1$ hold. Note that the trace itself is a similar set, containing all the positions where the atoms (with a height 1) hold. Based on the set it receives, the $l$-th iteration can then calculate all the positions where the subformulae of height $l$ hold. Each iteration consists of three phases: 1) *read phase* that reads and splits the input; 2) *map phase* that associates each formula with its superformula; and 3) *reduce phase* that applies the semantics of the appropriate subformula of $\Phi$. The final set of tuples represents all the positions where the input MTL formula holds, thus producing the verdict is only a matter of checking if it holds in the first position.

**Read phase.** The input reader component of the MapReduce framework is used in this phase; this component can process the input trace in a parallel way. The trace saved in a distributed file system is split into several blocks (usually 64MB in size), replicated (usually 3 times) and distributed evenly among the nodes. The MapReduce framework exploits this block-level parallelization both during the read and map phases. For example, the default block size of the Hadoop deployment is 64MB, which means that a 1GB trace is split in 16 parts and can be potentially processed using 16 parallel readers and mappers. However, if we execute the algorithm on 3 nodes with 4 cores each, we could process up to 12 blocks in parallel. The input reader is used only in the first iteration and can be seen as a parser that converts the trace into a uniform internal representation that is used in the subsequent iterations. As shown in Figure 6a, the $k$-th instance of the input reader handles the $k$-th block $T_k$ of the trace $T$. For each element $(\sigma, \tau)$ in $T_k$ and every atom $p$ occurring in the MTL formula $\Phi$, the reader outputs a key-value pair of the form $(p, (p \in \sigma, \tau))$. The key is the atom $p$ itself, while the value is a pair consisting of the truth value of $p$ at time $\tau$ (obtained by evaluating the expression $p \in \sigma$) and the time-stamp $\tau$.

**Map phase.** Each tuple generated by an input reader is passed to a mapper on the same node. Mappers associate the formula in the tuple with all its superformulae in $\Phi$. For example, given $\Phi = (a \wedge b) \vee \neg a$, if the input reader returns a tuple $(a, (\top, 42))$, the mapper will associate it with formulae $a \wedge b$ and $\neg a$, outputting the tuples $(a \wedge b, (a, \top, 42))$ and $(\neg a, (a, \top, 42))$. The mapper, shown in Figure 6b, receives tuples in the form $(\phi, (v, \tau))$ from the input reader and outputs all tuples of the form $(\psi, (\phi, v, \tau))$ where $\psi \in \sup_\Phi(\phi)$.

To support lazy semantics, the algorithm needs to consider all the positions in the trace where we want to evaluate the temporal operators. If any of these positions did not exist in the trace then the original algorithm would evaluate a formula to false (see the example in Section 4). However, to support lazy semantics, we do not need to introduce a position in the trace for each time instant: we know a priori that only formulae of the form $F_{=K}$ —explicitly introduced

```
1: function INPUT READER_{Φ,l}(T_k[])
2:   for all (σ, τ) ∈ T_k[] do
3:     for all p ∈ sub_a(Φ) do
4:       output(p, (p ∈ σ, τ))
5:     end for
6:   end for
7: end function
```
**(a)** Input reader algorithm

```
1: function MAPPER_{Φ,K,l}((φ, (v, τ)))
2:   for all ψ ∈ sup_Φ(φ) do
3:     output(ψ, (φ, v, τ))
      if lazy(ψ) then
        output(ψ, (φ_act, ⊥, τ + K))
      end if
4:   end for
5: end function
```
**(b)** Mapper algorithm

```
1: function REDUCER^{F_I}_{Φ,l}(ψ, T[])
2:   val ← ⊥, win ← ∅
3:   for all (φ, v, τ) ∈ checkDup(T) do
4:     win ← win ∪ (φ, v, τ) if (v)
5:     while ⌈win⌉_τ − ⌊win⌋_τ ∉ 0 ⊎ I do
6:       win ← win \ argmax_τ(win)
7:     end while
8:     val ← ∃τ' ∈ {win}_τ : τ' − τ ∈ I
9:     output(ψ, (val, τ))
10:  end for
11: end function
```
**(c)** Reducer for operator $F_I$

```
1: function REDUCER^{G_I}_{Φ,l}(ψ, T[])
2:   val ← ⊤, win ← ∅
3:   for all (φ, v, τ) ∈ checkDup(T) do
4:     win ← win ∪ (φ, v, τ) if (¬v)
5:     while ⌈win⌉_τ − ⌊win⌋_τ ∉ 0 ⊎ I do
6:       win ← win \ argmax_τ(win)
7:     end while
8:     val ← ∃τ' ∈ {win}_τ : τ' − τ ∈ I
9:     output(ψ, (¬val, τ))
10:  end for
11: end function
```
**(d)** Reducer for operator $G_I$

Figure 6: Reader, Mapper and Reducer algorithms. (Sets $\mathsf{sub}_a$ and $\mathsf{sup}_\Phi$ are defined in Section 6)

by the $\mathcal{L}_K$ decomposition— may be evaluated incorrectly if the appropriate positions are missing in the trace (see Example 1). Therefore, we modify the algorithm for the mapper (see Figure 6b) to introduce one position at $\tau + K$ only when the parent formula $\psi$ is a subformula of the form $F_{=K}$; this condition is captured by the *lazy()* predicate. The emitted tuple contains the tuple $(\varphi_{act}, \bot, \tau + K)$ as its value. The mapper is stateless and cannot check if a tuple at time instant $\tau + K$ already exists: it is the reducer's responsibility to discard a tuple if it has a duplicate.

**Reduce phase.** The reducers exploit the information produced by the mappers to determine the truth values of the superformula at each position, i.e., reducers apply the appropriate MTL semantics for the operator used in the superformula. The total number of reducers running in parallel at the $l$-th iteration is the minimum between the number of subformulae with height $l$ in the input formula $\Phi$ and the number of available reducers[4]. Each reducer calls an appropriate reduce function depending on the type of formula used as key in the received tuple. For space reasons we focus only on two algorithms: the one for the metric "*Eventually*" operator $F_I$ and the one for the metric "*Globally* operator $G_I$. We refer the reader to our previous work [9] for the full description of all the reducer algorithms.

Figure 6c shows the algorithm for formulae of the form $F_I \phi$. It uses an auxiliary boolean variable *val* and a queue *win*. The algorithm loops through all the tuples received in $T$, already sorted (by the *shuffle and sort* phase of the MapReduce framework) in descending order with respect to the time-stamps[5], and with all duplicates of tuple $(\varphi_{act}, \bot, \tau)$ discarded (by means of the *checkDup()* function). The queue *win* keeps track of all the tuples with positive truth value that fall in the convex union[6] (denoted as $\uplus$) of the intervals $[0, 0]$ and $I$. This is ensured by the inner while loop, which compares the minimal ($\lfloor win \rfloor_\tau$) and maximal ($\lceil win \rceil_\tau$) time-stamp in the queue and keeps removing the maximal tuple ($\mathsf{argmax}_\tau(win)$) until the loop condition is satisified. The final truth value of $\mathsf{F}_I\phi$ depends on whether the queue $win$ contains a tuple with a time-stamp $\tau'$ that is in the interval $I$. Notice that the size of the queue $win$ depends directly on the size of the interval $I$ and hence the memory scalability of the algorithm on individual nodes depends on the size of the intervals in formula $\Phi$.

---
[4]This depends on the configuration of the cluster. Typically, the number of reducers is the number of nodes in the cluster multiplied by the number of cores available on each node.
[5]Sorting intermediate tuples is called secondary sorting and for simplicity we omit the implementation details.
[6]A convex union of intervals is defined as a convex hull of the union of the intervals.

| $l$ | Atoms: | {$p$} | {$p$} |  | {$q$} |  | {$p,q$} |  | {$p,q$} | {$q$} |  | {$q$} |  |  |
|---|---|---|---|---|---|---|---|---|---|---|---|---|---|---|
|   | Time-Stamps: | 1 | 2 | 3 | 4 | 5 | 6 | 7 | 8 | 9 | 10 | 11 | 12 | 13 14 |
| 1 | $p$ | ⊤ | ⊤ |  | ⊥ |  | ⊤ |  | ⊤ | ⊥ | ⊥ |  |  |  |
|   | $\mathsf{F}_{[3,4]}(p)$ | ⊤ | ⊤ |  | ⊤ | ⊤ | ⊥ |  | ⊥ | ⊥ | ⊥ |  |  |  |
|   | $\mathsf{F}_{[0,3]}(p)$ | ⊤ | ⊤ |  | ⊤ | ⊤ | ⊤ |  | ⊤ | ⊥ | ⊥ |  |  |  |
| 2 | $\varphi_{act}$ |  |  |  |  | ⊥ | ⊥̸ |  | ⊥̸ |  | ⊥̸ | ⊥ | ⊥ | ⊥ |
|   | $\mathsf{F}_{[4,4]}(\mathsf{F}_{[0,3]}(p))$ | ⊤ | ⊤ |  | ⊤ | ⊥ | ⊥ |  | ⊥ | ⊥ | ⊥ | ⊥ | ⊥ | ⊥ |
| 3 | $\mathcal{L}_4(l2p(\Phi))$ | (⊤) | (⊤) |  | (⊤) | ⊤ | (⊥) |  | (⊥) | (⊥) | (⊥) | ⊥ | ⊥ | ⊥ |

**Figure 7: Evaluation of the $\mathcal{L}_4(l2p(\Phi)) = \mathsf{F}_{[3,4]}(p) \vee \mathsf{F}_{[4,4]}(\mathsf{F}_{[0,3]}(p))$ formula under MTL$_\mathsf{L}$ semantics.**

The reducer algorithm in Figure 6d implements the semantics of formulae of the form $\mathsf{G}_I\phi$. The code is similar to the one for the operator $\mathsf{F}_I$. The only difference is that the queue $win$ keeps track of all the tuples with negative truth value; hence, the truth value of $\mathsf{G}_I\phi$ depends on whether the queue $win$ contains a tuple in the interval $I$ that is a witness to the violation of $\mathsf{G}_I\phi$.

**Examples of application of the algorithm.**

Let us use our algorithm to evaluate the formula $\Phi$ from Example 1 on the same trace using MTL$_\mathsf{P}$ semantics. In the `read phase` the algorithm parses the trace in parallel and creates the input tuples for the `map phase`. From the first element ({$p$}, 1) the `InputReader` creates only the tuple $(p, (\top, 1))$ since $\Phi$ refers only to atom $p$. Tuples $(p, (\top, 1))$, $(p, (\top, 2))$, $(p, (\bot, 4))$, $(p, (\top, 6))$, $(p, (\top, 8))$, $(p, (\bot, 9))$, $(p, (\bot, 10))$ are thus received by the `map phase`. The `Mapper` associates the formulae from the input tuples with their superformulae. In the case of tuple $(p, (\top, 1))$ it generates only tuple $(\mathsf{F}_{[3,7]}(p), (p, \top, 1))$ since $\mathsf{F}_{[3,7]}(p)$ is the only superformula of $p$. The `Reducer phase`, therefore, receives tuples $(\mathsf{F}_{[3,7]}(p), (p, (\bot, 10)))$, $(\mathsf{F}_{[3,7]}(p), (p, (\bot, 9)))$, $(\mathsf{F}_{[3,7]}(p), (p, (\top, 8)))$, $(\mathsf{F}_{[3,7]}(p), (p, (\top, 6)))$, $(\mathsf{F}_{[3,7]}(p), (p, (\bot, 4)))$, $(\mathsf{F}_{[3,7]}(p), (p, (\top, 2)))$, $(\mathsf{F}_{[3,7]}(p), (p, (\top, 1)))$, all shuffled and sorted in a descending order of their time-stamps. Since all the tuples have the same key only one reducer is needed. The reducer applies the algorithm shown in Figure 6c and outputs the truth values of $\mathsf{F}_{[3,7]}(p)$ for every position in the trace: $(\mathsf{F}_{[3,7]}(p), (\bot, 10))$, $(\mathsf{F}_{[3,7]}(p), (\bot, 9))$, $(\mathsf{F}_{[3,7]}(p), (\bot, 8))$, $(\mathsf{F}_{[3,7]}(p), (\bot, 6))$, $(\mathsf{F}_{[3,7]}(p), (\top, 4))$, $(\mathsf{F}_{[3,7]}(p), (\top, 2))$, $(\mathsf{F}_{[3,7]}(p), (\top, 1))$. Notice that, the boolean values in the tuples correspond to the values in Figure 2 (row #4).

Assuming again that the memory requirement of keeping 8 positions is too demanding for our infrastructure we can now use parametric decomposition and lazy semantics to limit the upper bound of the interval in $\Phi$ to $K = 4$. We obtain formula $\mathcal{L}_4(l2p(\Phi)) = \mathsf{F}_{[3,4]}(p) \vee \mathsf{F}_{[4,4]}(\mathsf{F}_{[0,3]}(p))$.

Let us evaluate formula $\mathcal{L}_4(l2p(\Phi))$ on the same trace from Example 1 according to MTL$_\mathsf{L}$ semantics. Table 7 shows the truth values of the emitted tuples for every evaluated subformulae of $\mathcal{L}_4(l2p(\Phi))$. Since $h(\mathcal{L}_4(l2p(\Phi))) = 4$ the algorithm performs three iterations. The truth values of the subformulae from the different iterations are separated by the horizontal dashed lines. In the first iteration the trace is parsed to obtain the truth values of atom $p$. After that, two reducers in parallel calculate the truth values of the $\mathsf{F}_{[0,3]}(p)$ and $\mathsf{F}_{[3,4]}(p)$ subformulae. In the second iteration the Mapper emits the additional $\varphi_{act}$ tuples since the superformula is of the form $\mathsf{F}_{=4}$. The reducer evaluating formula $\mathsf{F}_{[4,4]}(\mathsf{F}_{[0,3]}(p))$ receives the tuples with the evaluation of $\mathsf{F}_{[0,3]}(p)$ and $\varphi_{act}$. The $\varphi_{act}$ tuples with the crossed truth values are discarded because since they are duplicates of the already existing $\mathsf{F}_{[0,3]}(p)$ tuple shown in the row above. Finally, in the third iteration we can see that the truth values $\mathcal{L}_4(l2p(\Phi))$ (circled in Figure 7) are the same (at all positions in common) as the truth values of $\Phi$ shown in Figure 2.

## 7. EVALUATION

We have implemented our trace checking algorithm in the MTLMAPREDUCE tool, which is publicly available [19]. The tool is implemented in Java and uses the Apache Spark framework [26], which supports iterative MapReduce applications in a better way than Apache Hadoop [2].

In this section we report on the evaluation of our tool, in terms of scalability and time/memory tradeoffs. More specifically, we evaluate our new trace checking algorithm by answering the following research questions:

RQ1: *How does the proposed algorithm scale with respect to the size of the time interval used in the formula to be checked?* (Section 7.2)

RQ2: *When compared to state-of-the-art tools, does the proposed algorithm have a better memory scalability with respect to the size of the time interval used in the formula to be checked?* (Section 7.2)

RQ3: *What are the time/memory tradeoffs of the proposed algorithm with respect to the decomposition parameter $K$?* (Section 7.3)

### 7.1 Evaluation settings

To evaluate our approach, we used six *t2.micro* instances from the Amazon EC2 cloud-based infrastructure with a single CPU core and 1 GB of memory each. We used the standard configuration for the HDFS distributed file system and the YARN data operating system. HDFS block size was set to 64 MB and block replication was set to 3. YARN was configured to allocate containers with memory between 512 MB and 1 GB with 1 core. In all the executions, we limited the memory of our algorithm to 1 GB.

Measuring the actual memory usage of user-defined code in Spark-based applications requires to distinguish between the memory usage of the Spark framework itself and the one of user-defined code. This step is necessary since the framework may use the available memory to cache intermediate data to speedup computation. Hence, to measure the memory usage of the auxiliary data structures used by our algorithm (e.g., the *win* queue), we instrumented the code. This instrumentation, which has a negligible overhead, monitors the memory usage of the algorithm's data structures and reports the maximum usage for each run.

For the evaluation described in the next two subsections, we used *synthesized* traces. By using synthesized traces, we are able to control in a systematic way the factors, such as the trace length and the frequency of events, that impact on the time and memory required for checking a specific type of formula. In particular, we evaluated our approach by triggering the worst-case scenario, in terms of memory scalability, for our trace checking algorithm. Such scenario is characterized by having the auxiliary data structures used by the algorithm always at their maximum capacity. To synthesize the traces, we implemented a trace generator program that takes as parameters the desired trace length $n$ and the number of events (i.e., atoms) $m$ per trace element. The program generates a trace with $n$ trace elements, such that the $i$-th element (with $0 \leq i \leq n-1$) has $i$ as time-stamp value. Each trace element has between 1 and $m$ events denoted as $\{p_1, \ldots, p_m\}$, where $p_1 = p$ and the other events are randomly selected with a uniform distribution from the set $\{p_2, \ldots, p_m\}$. We generated ten traces, with $n$ set to 50 000 000 and $m$ set to 20; the average size of each trace, before saving it in the distributed file system, is 3.2 GB. These traces and the other artifacts used for the evaluation are available on the tool web site [19].

## 7.2 Scalability

The performance of our distributed trace checking algorithm with respect to the length of the trace and the size of the formula has been already investigated in our previous work [9]. The same conclusions regarding these two parameters apply also to the new algorithm, which uses lazy semantics. Therefore, in this section we only focus on evaluating the memory scalability of the new algorithm.

To answer RQ1, we evaluate the memory usage of the algorithm for different sizes of the time interval used in the MTL formula to be checked. As discussed in Section 6, the largest time interval manageable in a cluster depends on the memory configuration of the node in the cluster with the least amount of memory available. Hence, we evaluate the memory usage on a single node by using formulae of height 1. We consider the two metric formulae $\mathsf{G}_{[0,N]}q$ and $\mathsf{F}_{[0,N]}p$, parametrized by the value $N$ of the bound of their time interval. Formula $\mathsf{F}_{[0,N]}p$ refers to atom $p$; notice that our trace generator guarantees that $p$ is present in every trace element. Formula $\mathsf{G}_{[0,N]}q$ refers to atom $q$; we configured our trace generator so that event $q$ is absent in all trace elements. These two formulae exercise the trace checking algorithm in its worst-case. Indeed, according to line 4 in Figure 6c, the reducer for $F_I$ buffers all the elements where atom $p$ is true; hence, when checking formula $\mathsf{F}_{[0,N]}p$, at any point in time the queue *win* will be at its maximum capacity. Dually, when checking formula $\mathsf{G}_{[0,N]}q$, the absence of the event $q$ from the trace will force the algorithm to maintain the queue *win* at its maximal capacity (line 4 in Figure 6d).

To answer RQ2, we need a baseline for comparison. Among the *non-distributed, non-parallel trace checking tools for* MTL, we selected the MONPOLY [6] tool, which was the best performing tool supporting MTL in the "offline monitoring" track of the first international Competition on Software for Runtime Verification [4] (CSRV 2014). However, it produced a stack overflow error when fed in input with the traces described above. Among distributed and parallel approaches, the only tool *supporting* MTL *and publicly available* is the one described in our previous work [9].

Plots in Figure 8a and Figure 8b show the execution time and the memory usage required to check, respectively, formula $\mathsf{G}_{[0,N]}q$ and $\mathsf{F}_{[0,N]}p$, instantiated with different values of parameter $N$. Each data point is obtained by running the algorithm over the ten synthesized traces and averaging the results. The plots colored in black show the average time and memory usage of our previous algorithm [9], which applies $\mathsf{MTL_P}$ semantics. The plots colored in gray represent the runs of our new algorithm that applies $\mathsf{MTL_L}$ semantics and decomposes all the formulae with time interval $N$ strictly greater than 30 000 000. The decomposition parameter $K = 30\,000\,000$ is the maximal value that our infrastructure can support.

Regarding RQ1, the gray plots confirm that the new algorithm can check, on very large traces, formulae that use very large time intervals in a reasonable time (at most 200s). As for RQ2, the algorithm from [9] exhausts the memory bound of 1GB for the evaluation of both formulae when the time interval $N$ is higher than 30 000 000. Conversely, our new algorithm uses at most 1GB of memory, showing a better memory scalability. Nevertheless, you can see that the new algorithm becomes about 1.5–1.8x slower that the previous algorithm when the time interval $N$ is higher than 30 000 000. This additional time is needed to process the new formula obtained through the $\mathcal{L}_K$ decomposition.

## 7.3 Time/memory tradeoffs

As suggested at the end of the previous section, the parametric decomposition used in the proposed trace checking algorithm leads to a reduced memory usage but increases the execution time. In this section we dig into and generalize this result by investigating the time/memory tradeoffs of our algorithm, with respect to the decomposition parameter $K$ (RQ3). More specifically, to answer RQ3 we evaluate the execution time and the memory usage of the algorithm for different values of parameter $K$, when checking formulae $\mathsf{G}_{[0,50\,000\,000]}q$ and $\mathsf{F}_{[0,50\,000\,000]}p$. These formulae are processed using the $\mathcal{L}_K$ decomposition, with values of $K$ that are taken from the set $V = \{\frac{5 \cdot 10^7}{i} \mid i = 2, 3, 4, \ldots\}$. As the set $V$ is potentially infinite, we set a threshold of one hour on the execution time.

The plots in Figure 8c show the execution time and the memory usage to check the two formulae. Each data point is obtained by running the algorithm over the ten synthesized traces and averaging the results. The value of $K$ is represented in both plots on the x-axis using the logarithmic scale. The smallest value of $K$ that satisfies the execution time threshold is 1 666 666 (obtained from set $V$ with $i = 30$); for this value of $K$ the algorithm used 54.14MB of memory and took 43 minutes to complete. As you can see from the plots, using a lower value for $K$ decreases the memory footprint of the algorithm. However, a lower value for $K$ also yields a longer execution time for the algorithm. This longer execution time is due to the fact that a lower value for $K$ increases the size (and the height) of the formula obtained after applying the $L_K$ decomposition. The increased height of the decomposed formula triggers more iterations of the algorithm, yielding longer execution times. There is clearly a tradeoff between time and memory determined by the value of the parameter $K$. A good balance is achieved when $K$ is set to the largest possible value supported by the infrastructure: in this way, it is possible to reduce the size of the decomposed formula without incurring a longer execution time for the algorithm. However, our algorithm is completely parametric in $K$, allowing engineers to tune the algorithm to be either more time- or more memory-intensive.

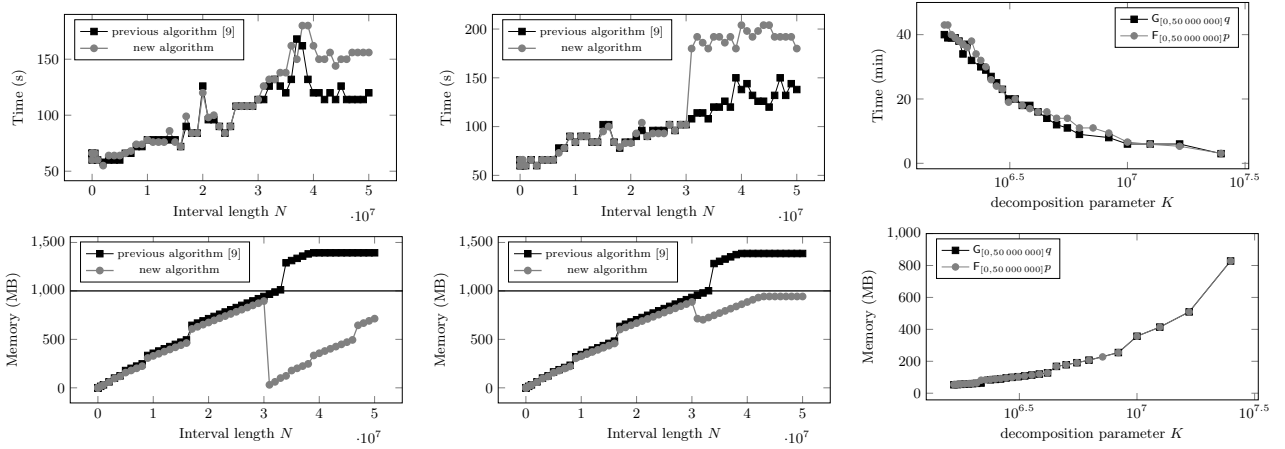

**(a)** Execution time (top) and memory usage (bottom) to check $\mathsf{G}_{=N}\,q$

**(b)** Execution time (top) and memory usage (bottom) to check $\mathsf{F}_{=N}\,p$

**(c)** Execution time (top) and memory usage (bottom) when varying parameter $K$

**Figure 8:** Scalability and time/memory tradeoffs for the proposed trace checking algorithm.

## 8. RELATED WORK

The approach presented in this paper is strictly related to work done in the areas of alternative semantics for metric temporal logics and of trace checking/run-time verification.

**Alternative semantics for metric temporal logics.** The work closest to our lazy semantics is the one in [14], which proposes an alternative MTL semantics, used to prove that signal-based semantics is more expressive than point-based semantics over finite words. Despite the similarity between the two semantics, the definition of the *Until* operator under our lazy semantics is more practical for the purpose of trace checking, since it requires the left subformula of an *Until* operator to hold in a finite number of positions. Reference [12] presents a revised semantics for the model parametric semantics (MPS) on finite domains of the TRIO temporal logic [21], to enable the executability of temporal logic specifications. The revised semantics addresses the limitations of the original MPS semantics when dealing with formulae with bounded temporal operator. It proposes an interpretation of bounded temporal operators that shares the same intuition behind the definition of lazy semantics.

**Trace checking/run-time verification.** Several approaches for trace checking and run-time verification and monitoring of temporal logic specifications have been proposed in the last decade. The majority of them (see, for example, [7, 15, 17, 23, 24]) are centralized and use sequential algorithms to process the trace (or, in online algorithms, the stream of events). As mentioned in Section 7.2, the centralized, sequential nature of these algorithms does not allow them either to process large traces or properties containing very large time bounds. In the last years there have been approaches for trace checking [5] and runtime verification [8, 20, 23] that rely on some sort of parallelization. However, they mostly focus on splitting the traces based on the data they contain, rather than on the structure of the formula. These approaches adopt first-order relations with finite domains to represent the events in the trace. The trace can then be split into several unrelated partitions based on the terms occurring in the relations. We consider these approaches *orthogonal* to ours, since we focus on the scalability with respect to the temporal dimension, rather than the data dimension. As for the specific application of MapReduce for trace checking, an iterative algorithm for LTL is proposed in [3]. Similarly to the algorithm presented in this paper and to our previous work [9], the algorithm in [3] performs iterations of MapReduce jobs depending on the height of the formula to check. However, it does not address the issue of memory consumption of the reducers. Moreover, the whole trace is kept in memory during the reduce phase, making the approach unfeasibile for very large traces.

## 9. CONCLUSIONS AND FUTURE WORK

The goal of this work is to address the memory scalability issue that affects trace checking algorithms when dealing with temporal properties that use large time intervals. We have proposed an alternative, *lazy* semantics for MTL, whose properties allow for a parametric decomposition of any MTL formula into an equivalent MTL formula with bounded time intervals. As shown in the evaluation, such decomposition can be used to improve distributed trace checking algorithms, making them more memory-efficient and able to deal with both very large traces and very large time intervals.

One future direction is to extend lazy semantics to a version of MTL with support for first-order relations on finite domains, to support more expressive properties. Another line of future research will focus on techniques to automate the $\mathcal{L}_K$ decomposition of formulae, for example to automatically determine the most appropriate value for $K$ based on the configuration of the available cloud infrastructure.

# APPENDIX

## A. PROOF OF THE LEMMATA

LEMMA 1. *Given an MTL formula $\phi$ and a timed word $\omega = (\sigma, \tau)$, for any $i \geq 0$, the following equivalence holds:*

$$(\sigma, \tau, i) \models_P \phi \text{ iff } (\sigma, \tau, \tau_i) \models_L l2p(\phi) \quad (1)$$

PROOF. The lemma is proved by structural induction on formula $\phi$. Let $\gamma$ be an MTL formula. The inductive hypothesis is $(\sigma, \tau, i) \models_P \gamma$ iff $(\sigma, \tau, \tau_i) \models_L l2p(\gamma)$.

1. base case $\gamma$ is $p \in \Pi$.
   $(\sigma, \tau, i) \models_P p$ iff $p \in \sigma_i$ iff $\exists i.(0 \leq i < |\sigma| \wedge \tau_i = \tau_i \wedge p \in \sigma_i)$ iff $(\sigma, \tau, \tau_i) \models_L p$. Then, we obtain $(\sigma, \tau, \tau_i) \models_L l2p(p)$, by definition of $l2p$.

2. $\gamma$ is $\neg \phi$.
   $(\sigma, \tau, i) \models_P \neg \phi$ iff $(\sigma, \tau, i) \not\models_P \phi$ iff, by inductive hypothesis, $(\sigma, \tau, \tau_i) \not\models_L l2p(\phi)$ iff, by definition of $l2p$, $(\sigma, \tau, \tau_i) \not\models_L \neg l2p(\neg\phi)$ iff $(\sigma, \tau, \tau_i) \models_L l2p(\neg \phi)$.

3. $\gamma$ is $\phi \wedge \psi$.
   $(\sigma, \tau, i) \models_P \phi \wedge \psi$ iff $(\sigma, \tau, i) \models_P \phi$ and $(\sigma, \tau, i) \models_P \psi$ iff, by inductive hypothesis, $(\sigma, \tau, \tau_i) \models_L l2p(\phi)$ and $(\sigma, \tau, \tau_i) \models_L l2p(\psi)$ iff, $(\sigma, \tau, \tau_i) \models_L l2p(\phi) \wedge l2p(\psi)$ iff, by definition of $l2p$, $(\sigma, \tau, \tau_i) \models_L l2p(\phi \wedge \phi)$.

4. $\gamma$ is $\phi U_I \psi$.
   $(\sigma, \tau, i) \models_P \phi U_I \psi$ iff
   $\exists j.(i \leq j < |\sigma| \wedge \tau_j - \tau_i \in I \wedge (\sigma, \tau, j) \models_P \psi \wedge \forall k.(i < k < j \to (\sigma, \tau, k) \models_P \phi)$ iff, by inductive hypothesis,
   $\exists j.(i \leq j < |\sigma| \wedge \tau_j - \tau_i \in I \wedge (\sigma, \tau, \tau_j) \models_L l2h(\psi) \wedge \forall k.(i < k < j \to (\sigma, \tau, \tau_k) \models_L l2p(\phi))$ iff
   $\exists t_j.(t_i \leq t_j < t_{|\sigma|} \wedge t_j - t_i \in I \wedge (\sigma, \tau, t_j) \models_L (\varphi_{act} \wedge l2h(\psi)) \wedge \forall t_k.(t_i < t_k < t_j \wedge \exists i.(0 \leq i < |\sigma| \wedge t_k = \tau_i) \to (\sigma, \tau, t_k) \models_L l2p(\phi))$ iff
   $(\sigma, \tau, t_i) \models_L l2p(\phi) U_I (\varphi_{act} \wedge l2p(\psi))$ iff, by definition of $l2p$
   $(\sigma, \tau, t_i) \models_L l2p(\psi U_I \psi)$.

The lemma is proved by considering $\gamma = \phi$. □

LEMMA 2. *For any timed word $(\sigma, \tau)$ and $t \geq 0$, we have that*

$$(\sigma, \tau, t) \models_L F_I F_J \phi \text{ iff } (\sigma, \tau, t) \models_L F_{I \oplus J} \phi. \quad (2)$$

PROOF. We show two cases. In the first one, both $F$ formulae have left and right-closed intervals. The second one considers all the other combinations.

$F_{[a,b]}(F_{[c,d]}\phi)$:
$(\sigma, \tau, t) \models_L F_{[a,b]}(F_{[c,d]}\phi)$ iff
$\exists t'.(t' \geq t \wedge t' - t \in [a,b] \wedge (\sigma, \tau, t') \models_L F_{[c,d]}\phi$ iff
$\exists t'.(t' \geq t \wedge t' \in [t+a, t+b] \wedge (\sigma, \tau, t') \models_L \exists t''.(t'' \geq t' \wedge t'' \in [t'+c, t'+d] \wedge (\sigma, \tau, t'') \models_L \phi)$ iff
$\exists t'.(t' \geq t \wedge (\sigma, \tau, t') \models_L \exists t''.(t'' \geq t' \wedge t'' \in [t+a+c, t+b+d] \wedge (\sigma, \tau, t'') \models_L \phi))$ iff
$\exists t''.(t'' \geq t \wedge t'' \in [t+a+c, t+b+d] \wedge (\sigma, \tau, t'') \models_L \phi$ iff
$(\sigma, \tau, t) \models_L F_{I \oplus J}\phi$.

$F_{\langle a,b \rangle}(F_{\langle c,d \rangle}\phi)$:
$(\sigma, \tau, t) \models_L F_{\langle a,b \rangle}(F_{\langle c,d \rangle}\phi)$ iff
$\exists t'.(t' \geq t \wedge t' - t \in \langle a,b \rangle \wedge (\sigma, \tau, t') \models_L F_{\langle c,d \rangle}\phi$ iff
$\exists t'.(t' \geq t \wedge t' \in \langle t+a, t+b \rangle \wedge (\sigma, \tau, t') \models_L \exists t''.(t'' \geq t' \wedge t'' \in \langle t'+c, t'+d \rangle \wedge (\sigma, \tau, t'') \models_L \phi)$ iff
$\exists t'.(t' \geq t \wedge (\sigma, \tau, t') \models_L \exists t''.(t'' \geq t' \wedge t'' \in (t+a+c, t+b+d) \wedge (\sigma, \tau, t'') \models_L \phi))$ iff
$\exists t''.(t'' \geq t \wedge t'' \in (t+a+c, t+b+d) \wedge (\sigma, \tau, t'') \models_L \phi$ iff
$(\sigma, \tau, t) \models_L F_{I \oplus J}\phi$.

□

LEMMA 3. *For any timed word $(\sigma, \tau)$ and $t \geq 0$, we have that*

$$(\sigma, \tau, t) \models_L F_I \phi \vee F_J \phi \text{ iff } (\sigma, \tau, t) \models_L F_{I \cup J}\phi, \text{ if } I \cap J \neq \emptyset.$$

PROOF. We prove the lemma for the case of $I = (a, b)$, $J = (c, d)$, as we can always rewrite intervals as left-right open ones. The case $F_{[0,b)}\phi$ becomes $F_{(1,b)}\phi \vee \phi$. The case for unbounded intervals is similar. By $I \cap J \neq \emptyset$ we have both $c + 1 < b$ and $a + 1 < d$ which entails $c < b$ and $a < d$. Therefore, we have that $\min\{a,b,c,d\} = \min\{a,c\}$ and $\max\{a,b,c,d\} = \max\{b,d\}$.
$(\sigma, \tau, t) \models_L F_I \phi \vee F_J \psi$ iff
$(\sigma, \tau, t) \models_L F_I \phi$ or $(\sigma, \tau, t) \models_L F_J \phi$ iff
$\exists t'.(t' \geq t \wedge t' - t \in (a,b) \wedge (\sigma, \tau, t') \models_L F_{(a,b)}\phi$ or $\exists t'.(t' \geq t \wedge t' - t \in (c,d) \wedge (\sigma, \tau, t') \models_L F_{(c,d)}\phi$ iff,
$\exists t'.(t' \geq t \wedge (t' - t \in (a,b) \wedge (\sigma, \tau, t') \models_L \phi \vee t' - t \in (c,d) \wedge (\sigma, \tau, t') \models_L \phi$ iff, as $a < b$ and $c < d$,
$\exists t'.(t' \geq t \wedge t' - t \in (\min\{a,c\}, \min\{b,d\}) \wedge (\sigma, \tau, t') \models_L \phi$ iff
$(\sigma, \tau, t) \models_L F_{(\min\{a,c\}, \max\{b,d\})}\phi$ iff
$(\sigma, \tau, t) \models_L F_{I \cup J}\phi$. □

## B. PROOF OF THE THEOREM 3

**Theorem 3** Given an $MTL_L$ formula $\phi$, timed word $(\sigma, \tau)$ and a positive constant $K$, the following equivalence holds:

$$(\sigma, \tau, 0) \models_L \phi \text{ iff } (\sigma, \tau, 0) \models_L \mathcal{L}_K(\phi) \quad (3)$$

and the right-hand side bound of every bounded interval in all temporal subformulae of $\mathcal{L}_K(\phi)$ is less than or equal to $K$.

PROOF. We can prove this statement by showing that $\mathcal{L}_K(\phi)$ can always be rewritten back to $\phi$ using lemmata 2 and 3. Let us preform structural induction on the $MTL_L$ formula $\phi$. The inductive hypothesis is $(\sigma, \tau, i) \models_L \theta$ iff $(\sigma, \tau, i) \models_L \mathcal{L}_K(\theta)$. Then, the theorem is proved by choosing $\theta = \phi$ and $i = 0$. In the proof we extensively use the following properties $\lfloor \frac{a}{K} + 1 \rfloor \cdot K = \lfloor \frac{a}{K} \rfloor \cdot K + K$ denoted with (*); $b - \lfloor \frac{b}{K} \rfloor \cdot K = b \mod K$ denoted with (**); and $\lfloor n + \epsilon \rfloor = n$, for $n \in \mathbb{N}$ and $\epsilon \in [0, 1)$ denoted with (***).

1. Base cases are the atoms which are not affected by the translation.
2. Same holds for boolean connectives.
3. Let $\theta = F_{\langle a,b \rangle}(\phi)$. We need to consider three cases.
   (a) $[b \leq K]$: $(\sigma, \tau, i) \models_L F_{\langle a,b \rangle}(\phi)$ iff $(\sigma, \tau, i) \models_L \exists j.(j - i \in \langle a,b \rangle$ and $(\sigma, \tau, j) \models_L \phi)$) which is, by inductive hypothesis, $(\sigma, \tau, i) \models_L \exists j.(j - i \in \langle a,b \rangle$ and $(\sigma, \tau, j) \models_L \mathcal{L}_K(\phi))$ iff $(\sigma, \tau, i) \models_L F_{\langle a,b \rangle}(\mathcal{L}_K(\phi))$ which is, by definition of $\mathcal{L}_K$, $(\sigma, \tau, i) \models_L \mathcal{L}_K(F_{\langle a,b \rangle}(\phi))$. Since $b \leq K$ the right-hand side bound is less then or equal to $K$.
   (b) $[K < b \leq \lfloor \frac{a}{K} + 1 \rfloor \cdot K]$: Identically to (a), we have $(\sigma, \tau, i) \models_L F_{\langle a,b \rangle}(\phi)$ iff $(\sigma, \tau, i) \models_L F_{\langle a,b \rangle}(\mathcal{L}_K(\phi))$ by inductive hypothesis. The interval is not bounded by $K$ as $K < b$. By property (**), we get $(\sigma, \tau, i) \models_L F_{\langle a \mod K + K \cdot \lfloor \frac{a}{K} \rfloor, b - K \cdot \lfloor \frac{a}{K} \rfloor + K \cdot \lfloor \frac{a}{K} \rfloor \rangle} \mathcal{L}_K(\phi)$ and, by Lemma 2, we obtain $(\sigma, \tau, i) \models_L F_{=K \cdot \lfloor \frac{a}{K} \rfloor}(F_{\langle a \mod K, b - \lfloor \frac{a}{K} \rfloor \cdot K \rangle} \mathcal{L}_K(\phi))$. By Corollary 1 the formula can be rewritten into

$(\sigma, \tau, i) \models_L \mathsf{F}_{=K}^{\lfloor \frac{a}{K} \rfloor}(\mathsf{F}_{\langle a \mod K, b - \lfloor \frac{a}{K} \rfloor \cdot K \rangle} \mathcal{L}_K(\phi))$ and, then, by definition of $\mathcal{L}_K$ we obtain $(\sigma, \tau, i) \models_L \mathcal{L}_K(\phi)$. By property (*) and the case assumption is $b \leq \lfloor \frac{a}{K} + 1 \rfloor \cdot K$ we have that $b - \lfloor \frac{a}{K} \rfloor \cdot K \leq K$ therefore the right-hand side bound of the interval is less than or equal to $K$.

(c) $[b > \lfloor \frac{a}{K} + 1 \rfloor \cdot K]$: Identically to (b) we have
$(\sigma, \tau, i) \models_L \mathsf{F}_{\langle a, b \rangle}(\phi)$ iff $(\sigma, \tau, i) \models_L \mathsf{F}_{=K}^{\lfloor \frac{a}{K} \rfloor}(\mathsf{F}_{\langle a \mod K, b - \lfloor \frac{a}{K} \rfloor \cdot K \rangle} \mathcal{L}_K(\phi))$.
Since $b > \lfloor \frac{a}{K} + 1 \rfloor \cdot K$ then $b - \lfloor \frac{a}{K} \rfloor \cdot K > K$. Let $n = \lfloor \frac{b}{K} \rfloor - \lfloor \frac{a}{K} \rfloor$, we can use Lemma 3 to write
$(\sigma, \tau, i) \models_L \mathsf{F}_{=K}^{\lfloor \frac{a}{K} \rfloor}(\mathsf{F}_{\langle a \mod K, K]}(\phi) \vee \mathsf{F}_{[K, 2K]} \mathcal{L}_K(\phi) \vee \mathsf{F}_{[2K, 3K]} \mathcal{L}_K(\phi) \vee \ldots \vee \mathsf{F}_{[(n-1)K, nK]} \mathcal{L}_K(\phi) \vee \mathsf{F}_{[nK, b - \lfloor \frac{a}{K} \rfloor \cdot K \rangle} \mathcal{L}_K(\phi))$
then, by Lemma 2 and property (*), we get
$(\sigma, \tau, i) \models_L \mathsf{F}_{=K}^{\lfloor \frac{a}{K} \rfloor}(\mathsf{F}_{\langle a \mod K, K]} \vee \mathsf{F}_{=K}(\mathsf{F}_{[0, K]} \mathcal{L}_K(\phi) \vee \mathsf{F}_{[K, 2K]} \mathcal{L}_K(\phi) \vee \ldots \vee \mathsf{F}_{[(n-2)K, (n-1)K]} \mathcal{L}_K(\phi) \vee \mathsf{F}_{[(n-1)K, b - \lfloor \frac{a}{K} + 1 \rfloor \cdot K \rangle} \mathcal{L}_K(\phi)))$.
The Lemma 2 is applied $n$ times until we get
$(\sigma, \tau, i) \models_L \mathsf{F}_{=K}^{\lfloor \frac{a}{K} \rfloor}(\mathsf{F}_{\langle a \mod K, K]} \vee \mathsf{F}_{=K}(\mathsf{F}_{[0, K]} \mathcal{L}_K(\phi) \vee \mathsf{F}_{=K}(\mathsf{F}_{[0, K]} \mathcal{L}_K(\phi) \vee \ldots \vee \mathsf{F}_{=K}(\mathsf{F}_{[0, K]} \mathcal{L}_K(\phi) \vee \mathsf{F}_{=K}(\mathsf{F}_{[0, b - \lfloor \frac{a}{K} + n \rfloor \cdot K \rangle} \mathcal{L}_K(\phi))) \ldots)))$.
According to properties (**) and (***) the value $b - \lfloor \frac{a}{K} + n \rfloor \cdot K = b \mod K$, which is strictly less than $K$.

By definition of $\mathcal{D}_F$ (base case) we write
$(\sigma, \tau, i) \models_L \mathsf{F}_{=K}^{\lfloor \frac{a}{K} \rfloor}(\mathsf{F}_{\langle a \mod K, K]} \vee \mathsf{F}_{=K}(\mathsf{F}_{[0, K]} \mathcal{L}_K(\phi) \vee \mathsf{F}_{=K}(\mathsf{F}_{[0, K]} \mathcal{L}_K(\phi) \vee \ldots \vee \mathsf{F}_{=K}(\mathsf{F}_{[0, K]} \mathcal{L}_K(\phi) \vee \mathsf{F}_{=K}(\mathcal{D}_F(\mathcal{L}_K(\phi), K, b - \lfloor \frac{a}{K} + n \rfloor \cdot K))) \ldots)))$.
By definition of $\mathcal{D}_F$ (recursive case) we write
$(\sigma, \tau, i) \models_L \mathsf{F}_{=K}^{\lfloor \frac{a}{K} \rfloor}(\mathsf{F}_{\langle a \mod K, K]} \vee \mathsf{F}_{=K}(\mathsf{F}_{[0, K]} \mathcal{L}_K(\phi) \vee \mathsf{F}_{=K}(\mathsf{F}_{[0, K]} \mathcal{L}_K(\phi) \vee \ldots \vee \mathsf{F}_{=K}(\mathcal{D}_F(\mathcal{L}_K(\phi), K, b - \lfloor \frac{a}{K} + n \rfloor \cdot K + K)) \ldots)))$.
By property (*) we write
$(\sigma, \tau, i) \models_L \mathsf{F}_{=K}^{\lfloor \frac{a}{K} \rfloor}(\mathsf{F}_{\langle a \mod K, K]} \vee \mathsf{F}_{=K}(\mathsf{F}_{[0, K]} \mathcal{L}_K(\phi) \vee \mathsf{F}_{=K}(\mathsf{F}_{[0, K]} \mathcal{L}_K(\phi) \vee \ldots \vee \mathsf{F}_{=K}(\mathcal{D}_F(\mathcal{L}_K(\phi), K, b - \lfloor \frac{a}{K} + n - 1 \rfloor \cdot K)) \ldots)))$.
We apply definition of $\mathcal{D}_F$ (recursive case) and property (*) $n-1$ times to get
$(\sigma, \tau, i) \models_L \mathsf{F}_{=K}^{\lfloor \frac{a}{K} \rfloor}(\mathsf{F}_{\langle a \mod K, K]} \vee \mathsf{F}_{=K}(\mathcal{D}_F(\mathcal{L}_K(\phi), K, b - \lfloor \frac{a}{K} + 1 \rfloor \cdot K)))$.
Finally, we apply the definition of $\mathcal{L}_K$ to obtain
$(\sigma, \tau, i) \models_L \mathcal{L}_K(\mathsf{F}_{\langle a, b \rangle}(\phi))$.

☐